\documentclass[conference]{IEEEtran}
\IEEEoverridecommandlockouts
\usepackage{cite}
\usepackage{amsmath,amssymb,amsfonts}
\usepackage{algorithmic}
\usepackage{graphicx}
\usepackage{textcomp}
\usepackage{xcolor}
\def\BibTeX{{\rm B\kern-.05em{\sc i\kern-.025em b}\kern-.08em
    T\kern-.1667em\lower.7ex\hbox{E}\kern-.125emX}}
\begin{document}

\title{Joint Source-Channel Coding for Wireless Image Transmission: A Deep Compressed-Sensing Based Method}

\author{\IEEEauthorblockN{Mohammad Amin Jarrahi}
\IEEEauthorblockA{School of Computer Science and \\Electronic Engineering (CSEE) \\
University of Essex\\
Colchester, United Kingdom \\
m.jarrahi@essex.ac.uk}
\and
\IEEEauthorblockN{Eirina Bourtsoulatze}
\IEEEauthorblockA{School of Computer Science and \\Electronic Engineering (CSEE) \\
University of Essex\\
Colchester, United Kingdom \\
e.bourtsoulatze@essex.ac.uk}
\and
\IEEEauthorblockN{Vahid Abolghasemi}
\IEEEauthorblockA{School of Computer Science and \\Electronic Engineering (CSEE) \\
University of Essex\\
Colchester, United Kingdom \\
v.abolghasemi@essex.ac.uk}
}

\maketitle

\begin{abstract}
Nowadays, the demand for image transmission over wireless networks has surged significantly. To meet the need for swift delivery of high-quality images through time-varying channels with limited bandwidth, the development of efficient transmission strategies and techniques for preserving image quality is of importance. This paper introduces an innovative approach to Joint Source-Channel Coding (JSCC) tailored for wireless image transmission. It capitalizes on the power of Compressed Sensing (CS) to achieve superior compression and resilience to channel noise. In this method, the process begins with the compression of images using a block-based CS technique implemented through a Convolutional Neural Network (CNN) structure. Subsequently, the images are encoded by directly mapping image blocks to complex-valued channel input symbols. Upon reception, the data is decoded to recover the channel-encoded information, effectively removing the noise introduced during transmission. To finalize the process, a novel CNN-based reconstruction network is employed to restore the original image from the channel-decoded data. The performance of the proposed method is assessed using the CIFAR-10 and Kodak datasets. The results illustrate a substantial improvement over existing JSCC frameworks when assessed in terms of metrics such as Peak Signal-to-Noise Ratio (PSNR) and Structural Similarity Index (SSIM) across various channel Signal-to-Noise Ratios (SNRs) and channel bandwidth values. These findings underscore the potential of harnessing CNN-based CS for the development of deep JSCC algorithms tailored for wireless image transmission.
\end{abstract}


\section{Introduction}
\subsection{Motivation}
Wireless transmission of images have faced various challenges in compression, transmission resilience, and quality preservation. Despite the popularity of the wireless image transmission systems, achieving reliable transfer with efficient image compression remains a challenge \cite{R1, R2}. Conventional approaches utilize separate source and channel coding methods. While this strategy has its merits, an alternative is needed to improve the performance in noisy and bandwidth-limited conditions \cite{R3}. The joint source-channel coding (JSCC) method offers a compelling alternative, integrating statistical image properties with channel characteristics to enhance compression efficiency and resilience to channel noise \cite{R4}.

Incorporating signal-processing concepts, particularly compressed sensing (CS), in the design of JSCC presents an opportunity to enhance wireless image transmission \cite{R5}. While CS has demonstrated the ability to recover sparse signals, its sparsity assumption and expensive reconstruction process motivate new efficient methods. Deep Learning (DL) provides a suitable option for CS to address its limitations and improve the efficiency \cite{R6}. This paper aims to leverage DL-based CS within the JSCC framework for wireless image transmission. Through DL-based sampling and advanced reconstruction, the goal is to propose a novel approach that enhances compression efficiency and preserves image quality in challenging conditions.

\begin{figure*}[!t]
	\centering
	\includegraphics[scale=0.9]{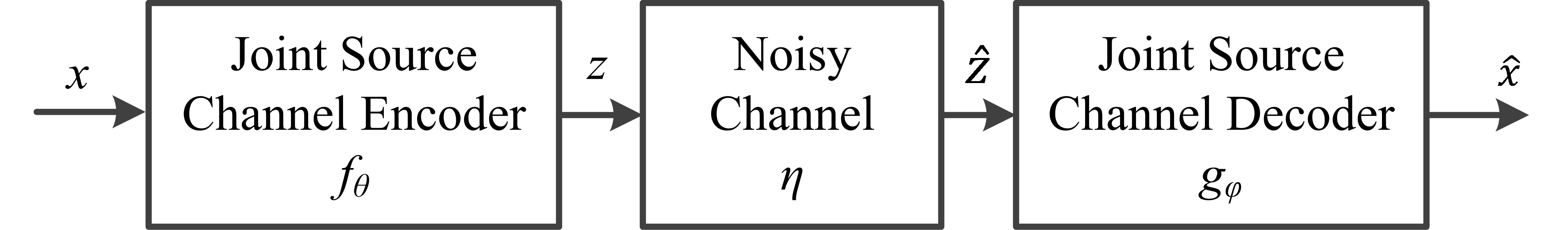}
	\caption {Components of the system model \cite{R7}}
	\vspace{-4mm}
\end{figure*}

\subsection{Literature Review}
Up to now, various techniques and frameworks have been proposed for JSCC and CS methods. In the following, these methods are categorized and some important relevant papers are reviewed and summarized. 
\subsubsection{DL-based JSCC for wireless image transmission}

DL-based methods have offered a new option for JSCC by leveraging encoder and decoder network models that learn from input images. These models are typically auto-encoders implemented as deep networks. In this methodology, the encoder output forms the transmitted code-word, which is a compressed representation of the source. The paired decoder at the receiver end aims to reconstruct the original source image by decoding the received noisy code-word, which is essentially a latent variables distorted by channel noise.
For instance, Bourtsoulatze et \textit{al.} introduce an end-to-end model that exhibits high performance, though it sacrifices interpretability \cite{R7}. Likewise, the authors in \cite{R8} present a DL-based JSCC approach capable of adapting to channel variations, which, however, suffers from a notable encoding delay.

\subsubsection{DL-based CS methods} 
DL excels in learning features for tasks like recognition and restoration, replacing traditional methods in CS. The DL models prioritize preserving image information, notably local features, by integrating stacked convolutional layers in the reconstruction process. This advancement significantly accelerates reconstruction speed compared to conventional methods. DL-based CS models dynamically adapt during training, eliminating the need for manual design. For example, Deep-CS \cite{R9} offers a straightforward CS approach, lacking robust theoretical guarantees. AMP-Net, a denoising-based approach with end-to-end training, aims to leverage prior knowledge but is sensitive to hyperparameter tuning \cite{R10}. Trans-CS, utilizing self-attention mechanisms, enhances reconstruction quality in compressed sensing \cite{R11}. Despite its flexibility, it is prone to overfitting.

\subsubsection{Application of CS in wireless image transmission systems}
Applying CS to wireless image transmission offers practical advantages, evident in solutions like SoftCast \cite{R12} and SparseCast \cite{R13}. SoftCast uses a Discrete Cosine Transform (DCT) on images and transmits coefficients directly through a dense constellation \cite{R12}. SparseCast encodes DCT coefficients, optimizing bandwidth with frequency domain sparsity and minimal metadata using fixed measurement levels \cite{R13}. While versatile, these methods are sensitive to channel changes. Song et al. propose a distributed CS for scalable cloud-based image transmission \cite{R14}. This strategy improves reconstruction using cloud resources, cutting transmission time and enhancing resistance to channel impairments. However, cloud disruptions may impact image quality and increase transmission errors.

\subsection{Contributions}
In this paper, we propose a novel JSCC algorithm that leverages the power of DL-based CS to achieve higher compression rate and better resilience to channel noise compared to state-of-the-art DL-based JSCC methods. In the proposed method, the images are firstly compressed and encoded using a novel CNN-based structure. This structure comprises a block-based CS (BCS) module realised via a convolutional neural network (CNN) which complements a DL-based source and channel encoder. The introduction of this module allows to leverage the properties of CS to enhance the performance of the DL-based JSCC scheme. The CS module captures the image's structural information which is then mapped to a complex-valued signal by the DL-based encoder. The compressed encoded images are then transmitted over a noisy channel modeled as a non-trainable layer. At the receiver side, a CNN-based decoder recovers the channel-encoded data and reconstructs the images. The decoder consists of a DL-based decoder network which recovers the channel-encoded information from the channel noise. This is then fed into a CNN-based reconstruction network, which reconstructs the original image from the compressed samples. The proposed JSCC algorithm leverages a DL-based sampling matrix and reconstruction capabilities  for improved image compression and reconstruction in wireless image transmission system. Numerical evaluations show that the proposed scheme significantly outperforms existing DL-based JSCC methods such as Deep JSCC (DJSCC) \cite{R7} and Attention DL based JSCC (ADJSCC) \cite{R15} with respect to various metrics. 

\section{System Model}
Consider a point-to-point image transmission system as shown in Fig. 1. An input image of size $H($height$) \times W($width$) \times C($number of channels$)$ is represented by a vector $x \in \mathbb{R}^n$, where $n = H \times W \times C$ and $\mathbb{R}$ denotes the set of real numbers.
The joint source-channel encoder encodes $x$ via the encoding function $f_\theta: \mathbb{R}^n \longrightarrow \mathbb{C}^k$ which produces a vector of complex-valued channel input symbols $z \in \mathbb{C}^k$. The encoding process can be expressed as:
\begin{equation}
z=f_\theta(x) \in \mathbb{C}^k
\end{equation}
where $k$ is the number of channel input symbols, $\theta$ is the parameter set of the joint source-channel encoder and $\mathbb{C}$ denotes the set of complex numbers. The encoder maps the $n$-dimensional vector of real-valued image $x$ to a $k$-dimensional vector of complex-valued channel input samples $z$.

To satisfy the average power constraint at the joint source-channel encoder, $\frac{1}{k} E\left(z z^*\right) \leq P$ is also imposed, where $z^*$ denotes the complex conjugate of $z$ and $P$ is the average power constraint \cite{R3}. The encoded symbols $z$ are transmitted over a noisy channel represented by the function $\eta: \mathbb{C}^k \rightarrow \mathbb{C}^k$. Additive white Gaussian noise (AWGN) is considered in our work. The channel output symbols $\hat{z} \in \mathbb{C}^k$ received by the joint source-channel decoder are expressed as: 
\begin{equation}
\hat{z}=\eta(z)=z+\omega
\end{equation}

\noindent where the vector $\omega \in \mathbb{C}^k$ consists of independent and identically distributed $(i.i.d)$ samples with the distribution $C N\left(0, \sigma^2 I\right)$. $\sigma^2$ is the average noise power and $CN(.,.)$ denotes a circularly symmetric complex Gaussian distribution. The proposed method can be extended to other channel models which can be represented by a differentiable transfer function. The joint source-channel decoder uses a decoding function $g_{\varphi}: \mathbb{C}^k \rightarrow \mathbb{R}^n$ to map $\hat{z}$ as follows:
\begin{equation}
\hat{x}=g_{\varphi}(\hat{z})=g_{\varphi}\left(\eta\left(f_\theta(x)\right)\right)
\end{equation}
where $\hat{x} \in \mathbb{R}^n$ is an estimation of the original image $x$, and $\varphi$ is the parameter set of the joint source-channel decoder. In this paper, the encoder $f_\theta$ and decoder $g_{\varphi}$ functions are modeled using a novel CNN structure, as presented in the following section.

\section{Deep CS-Based JSCC}
\subsection{Model Architecture}
The architectural details of the encoder and decoder networks, along with their constituent blocks, are shown in Fig. 2. The JSCC encoder comprises a BCS sampling network, followed by an array of further processing blocks, which collectively realise image compression and resilience to channel-induced noise. Considering that the input channel statistics are generally not known at the decoder, the initial step involves normalizing input images based on the maximum pixel value of 255, thereby restricting pixel values to the [0, 1] range. Subsequently, these normalized pixels feed into the sampling layer, which gathers  CS measurements. The sampling layer, employing BCS \cite{R16}, generates compressed image samples.

\begin{figure*}[!t]
	\centering
	\includegraphics[scale=1.2]{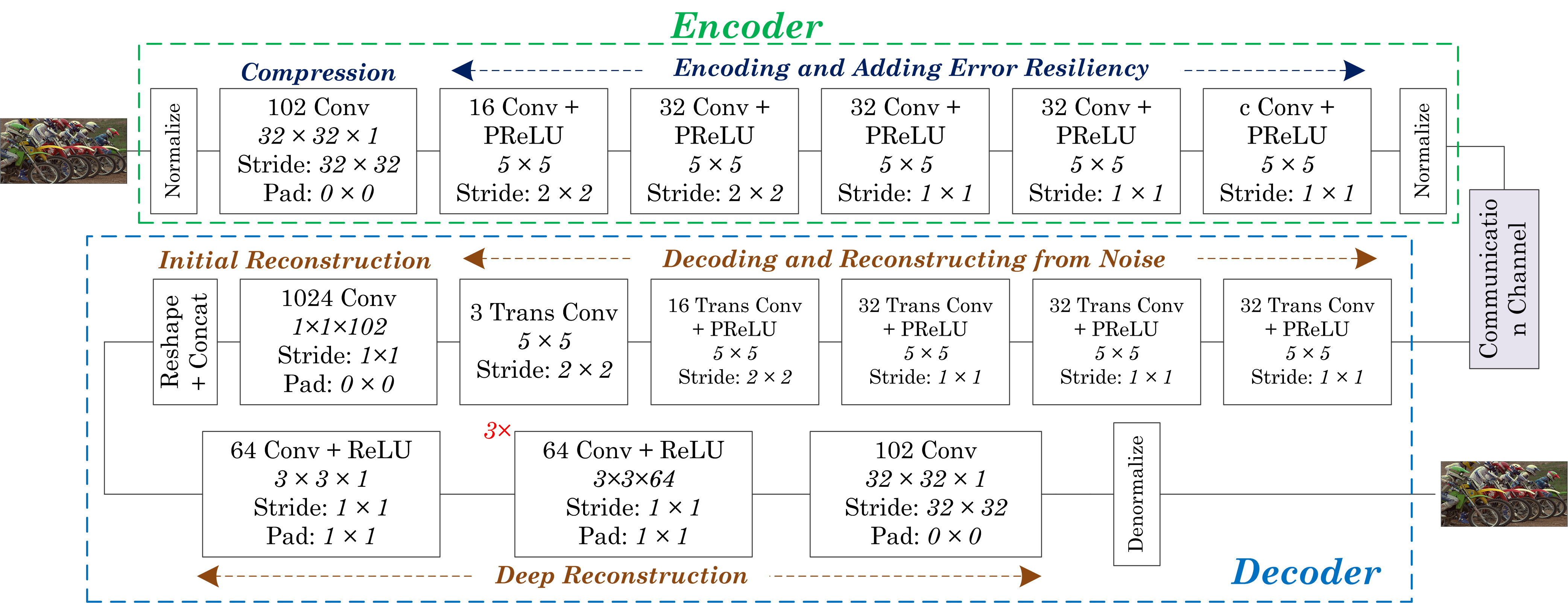}
	\caption {Architecture of the proposed model}
	\vspace{-4mm}
\end{figure*}

The CS sampling operates as follows \cite{R18}.  Initially, the image is partitioned into non-overlapping blocks, each having dimensions $B \times B \times l$. Here, $l$ denotes the number of colour channels, and $B$ signifies the block size. Compressed measurements are derived using a sampling matrix $\varphi_B$, with dimensions $n_B \times l B^2$. In scenarios where a sampling ratio, such as $L/V$, is applied, the dimensions of $\varphi_B$ become $(L/V) \times B$. The sampling process can be expressed as $y_j = \varphi_B x_j$, with $y_j$ and $x_j$ representing the $j^{th}$ block.
One important insight is that each row of the sampling matrix $\varphi_B$ can be perceived as a filter. Hence, a convolutional layer is adopted to simulate this compressed sampling process. Given that the size of every image block is $B \times B \times l$, the dimensions of each filter in the sampling layer are also $B \times B \times l$, allowing each filter to produce a single measurement. Notably, for non-overlapping sampling, the convolutional layer employs a stride of $B \times B$. There are no biases associated with these filters, and no activation functions are applied post-sampling. In essence, the output is $n_B$ feature maps, with each column of output encapsulating $n_B$ measurements originating from an image block. Importantly, the learning process encompasses learning the sampling matrix alongside other network parameters through an end-to-end optimization, as elaborated in subsequent sections.

Subsequent to the sampling layer, the data flow progresses through a sequence of convolutional layers, followed by the application of Parametric Rectified Linear Unit (PReLU) activation functions and a normalization layer. This sequence of convolutional layers takes on the role of extracting crucial features from the compressed image. These features are then amalgamated to generate the channel's input samples. The inclusion of nonlinear activation functions, represented here by PReLU, is pivotal. They facilitate the learning of a nonlinear mapping that maps the source signal space into the coded signal space. This is where the network can model complex, non-linear relationships within the data. As a final step within the encoder, the output of the last convolutional layer is subjected to a normalization process as follows:

\begin{equation}
z=\sqrt{k P} \frac{\tilde{z}}{\sqrt{\tilde{z}^* \tilde{z}}}
\end{equation}
where $\tilde{z}^*$ is the conjugate transpose of $\tilde{z}$, such that the channel input $z$ satisfies the average transmit power constraint $P$. 
It should be noted that the output of last CNN layer would be the input of normalization layer.
Following the encoding operation, the joint source-channel coded sequence is sent over the communication channel by directly transmitting the real and imaginary parts of the channel input samples over the I and Q components of the digital signal. The channel introduces random corruption to the transmitted symbols. To be able to optimize the proposed wireless image transmission system in an end-to-end manner, the communication channel must be incorporated into the overall architecture. Therefore, the communication channel is modeled as a non-trainable layer, which is represented by the transfer function in Eq. (2) [7]. 

The receiver comprises a joint source-channel decoder which reconstructs the received noisy compressed data. The decoder firstly maps the corrupted and compressed complex-valued signal to an estimation of the original channel input, then image blocks are reconstructed using a reconstruction network. Specifically, the decoder first inverts the operations performed by the encoder by passing the received corrupted coded inputs through a series of transpose convolutional layers with PReLU activation functions to map the image features to an estimate of the originally transmitted image block. 

The recovered CS measurements are then used to reconstruct the original image. The reconstruction network consists of an initial reconstruction network and a deep reconstruction network \cite{R18}. Similar to the compressed sampling process, a convolutional layer with appropriate kernel size and stride is utilized to implement the initial reconstruction process. $lB^2$ convolutional filters of size $1\times 1\times n_B$ are used to obtain each initial reconstructed block. Then, a combination layer, which contains a reshape function and a concatenation function, is utilized to obtain the initial reconstructed image. This layer first reshapes each $1\times 1\times B^2$ reconstructed vector to a $B\times B\times l$ block, then concatenates all blocks to get an initial reconstructed image. The initial reconstruction allows to reconstruct the entire image rather than an independent image block, thus making full use of both intra- and inter-block information for better reconstruction. Since there is no activation layer in the initial reconstruction network, it is a linear signal reconstruction network.

The initial reconstruction is followed by a non-linear reconstruction process which further improves the quality of the reconstructed image. In this paper, a deep sub-network is utilized \cite{R18}, called a deep reconstruction sub-network, which realises the non-linear reconstruction process. The deep reconstruction sub-network contains $m$ layers where the layers except the ﬁrst and the last are of the same type: $d$ filters of size $f\times f\times d$ where a ﬁlter operates on a $f\times f$ spatial region across $d$ channels (feature maps). The ﬁrst layer of the deep reconstruction sub-network operates on the initial reconstructed output, so that it has $d$ filters of size $f\times f\times 1$. The last layer, which outputs the ﬁnal image estimation, consists of a single ﬁlter of size $f\times f\times d$. In the experiments, $d$ and $f$ are set to $d=64$ and $f=3$. Furthermore, ReLU is also utilized as activation function after each convolution layer in the deep reconstruction sub-network.
\subsection{Loss Function}
The proposed encoder and the decoder network are optimized jointly in an end-to-end manner. Given the input image $x$, the goal is to obtain a highly compressed encoded measurement with the encoder, and then recover  the original input image $x$ from its noisy version with the decoder network. Since the encoder, decoder and communication channel form an end-to-end network, they can be trained jointly. Following most of DL based methods in this field, the mean square error is adopted as the cost function of the proposed network. The optimization objective is represented as:
\vspace{-0.2cm}
\begin{equation}
\min \frac{1}{ N} \sum_{i=1}^N\left\|g_{\varphi}\left(\eta\left(f_\theta(x_i)\right)\right)-x_i\right\|_2^2
\end{equation}
where $\varphi$ represents the parameter of the decoder network needed to be trained, and $g_{\varphi}\left(\eta\left(f_\theta(x_i)\right)\right)$ is the ﬁnal reconstructed output $\hat{x}$. Also, $N$ represents the number of samples or data points in the dataset. It should be noted that we train the encoder network and the decoder network jointly, but they can be utilized in the model independently.

\section{Results and Discussions}
The proposed model is implemented in Tensorflow and optimized using the Adam algorithm. The compression ratio $k/n$, defined as the ratio of the channel bandwidth $k$ to source bandwidth $n$, is changed from $0.05$ to $0.45$. Also, the channel signal-to-noise ratio (SNR), defined as:
\begin{equation}
S N R=10 \log _{10} \frac{P}{{\sigma}^2}(d B)
\end{equation}
is varied during different experiments. The performance of the algorithm is quantified in terms of peak SNR (PSNR) of the reconstructed images. PSNR is calculated as the ratio of the peak signal power ($Peak$) to the mean squared error ($MSE$) between the original and reconstructed images:
\begin{equation}
P S N R=10 \log _{10} \frac{P e a k^2}{M S E}(d B)
\end{equation}
We train our proposed JSCC architecture on both CIFAR-10 and Imagenet datasets and compare the results with state-of-the-art deep learning-based JSCC methods, namely DJSCC [7] and ADJSCC [15].
\subsection{Evaluation on CIFAR-10 Dataset }
The training dataset comprises $60000$ images, each sized $32\times 32\times 3$, alongside randomly generated realizations of the communication channel [7]. To gauge the effectiveness of the proposed technique, it is assessed on a distinct set of 10000 test images from the CIFAR-10 dataset, these images being separate from those used during training. Initially, a learning rate of $10^{-3}$ is employed, which is then lowered to $10^{-4}$ after 500000 iterations. Training is conducted using mini-batches, each containing $64$ samples, until the performance on the test dataset no longer shows improvement. The following values are considered in the experiments for this dataset: $B=8$ and $l=3$. It is important to note that the test set images are not employed for tuning network hyperparameters. To account for the influence of channel-induced randomness, each image is transmitted 10 times during performance evaluation. The study examines the performance of the proposed algorithm within AWGN environment, with SNR being adjusted to varying levels.

In Fig. 3, the performance of the proposed algorithm on CIFAR-10 test images with respect to the compression ratio, for different SNR values is shown. It must be noted that for each case, the same SNR value is used in training and evaluation. The results show that the proposed method significantly outperforms the state-of-the-art methods DJSCC and ADJSCC across the entire range of compression ratio and for both low and medium SNR values. 

\begin{figure}[!t]
	\centering
	\includegraphics[scale=0.6]{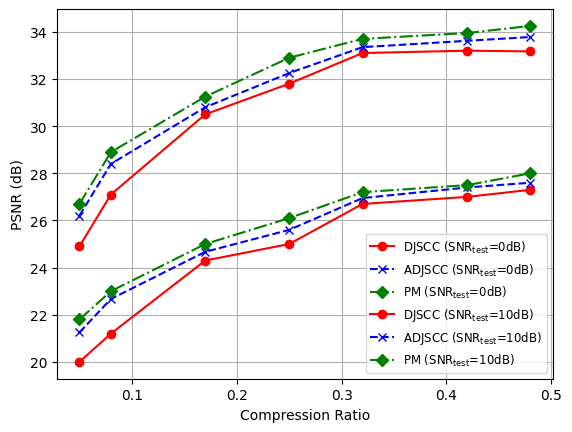}
	\caption {CIFAR-10 dataset: performance of the proposed method versus varying compression ratios over an AWGN channel (PM=Proposed Method)}
	\vspace{-4mm}
\end{figure}

\begin{figure}[!t]
	\centering
	\includegraphics[scale=0.6]{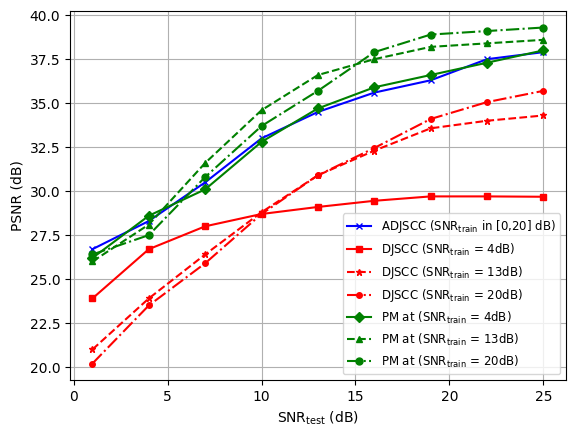}
	\caption {CIFAR-10 dataset: performance of different methods with compression ratio $1/6$, versus varying channel SNRs over an AWGN channel (PM=Proposed Method)}
	\vspace{-4mm}
\end{figure}

Fig. 4 depicts the PSNR of the reconstructed images against the SNR of the channel, with a fixed compression ratio of $1 / 6$. Each curve within Fig. 4 is generated by training the proposed end-to-end system at a specific channel SNR value, referred to as $SNR_{\text{train}}$. Subsequently, the performance of the learned encoder/decoder parameters is assessed using test images under various SNR conditions, designated as $SNR_{\text{test}}$. Essentially, each curve illustrates the effectiveness of the proposed approach when optimized for a channel SNR equal to $SNR_{\text{train}}$, then tested in distinct channel conditions corresponding to $SNR_{\text{test}}$. These findings shed light on the algorithm behavior when operating in channel conditions divergent from the optimization scenario, highlighting its resilience to channel quality variations. The outcomes highlight that the proposed method consistently outperforms trained DJSCC. Moreover, both the proposed method and ADJSCC exhibit adaptability to changing SNR, evident from their smooth decline in performance as SNR decreases. Notably, the proposed method holds an advantage over ADJSCC, showcasing superior performance as $SNR_{\text{test}}$ increases, surpassing ADJSCC ($SNR_{\text{train}} = 4\text{dB}$) by up to $1.5\text{dB}$.

\begin{figure}[!t]
	\centering
	\includegraphics[scale=0.6]{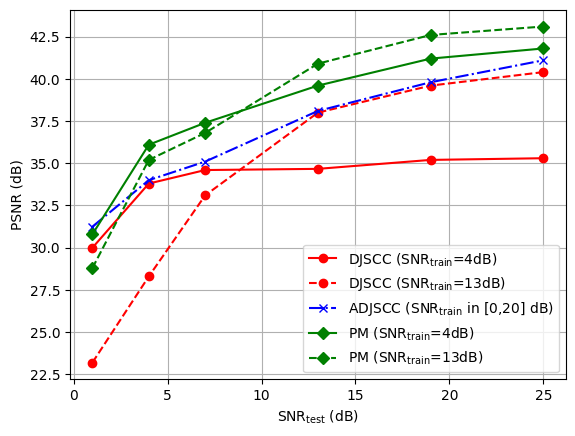}
	\caption {Kodak dataset: performance of different methods with compression ratio $1/6$, versus varying channel SNRs over an AWGN channel (PM=Proposed Method)}
	\vspace{-4mm}
\end{figure}

\begin{figure}[!t]
	\centering
	\includegraphics[scale=0.42]{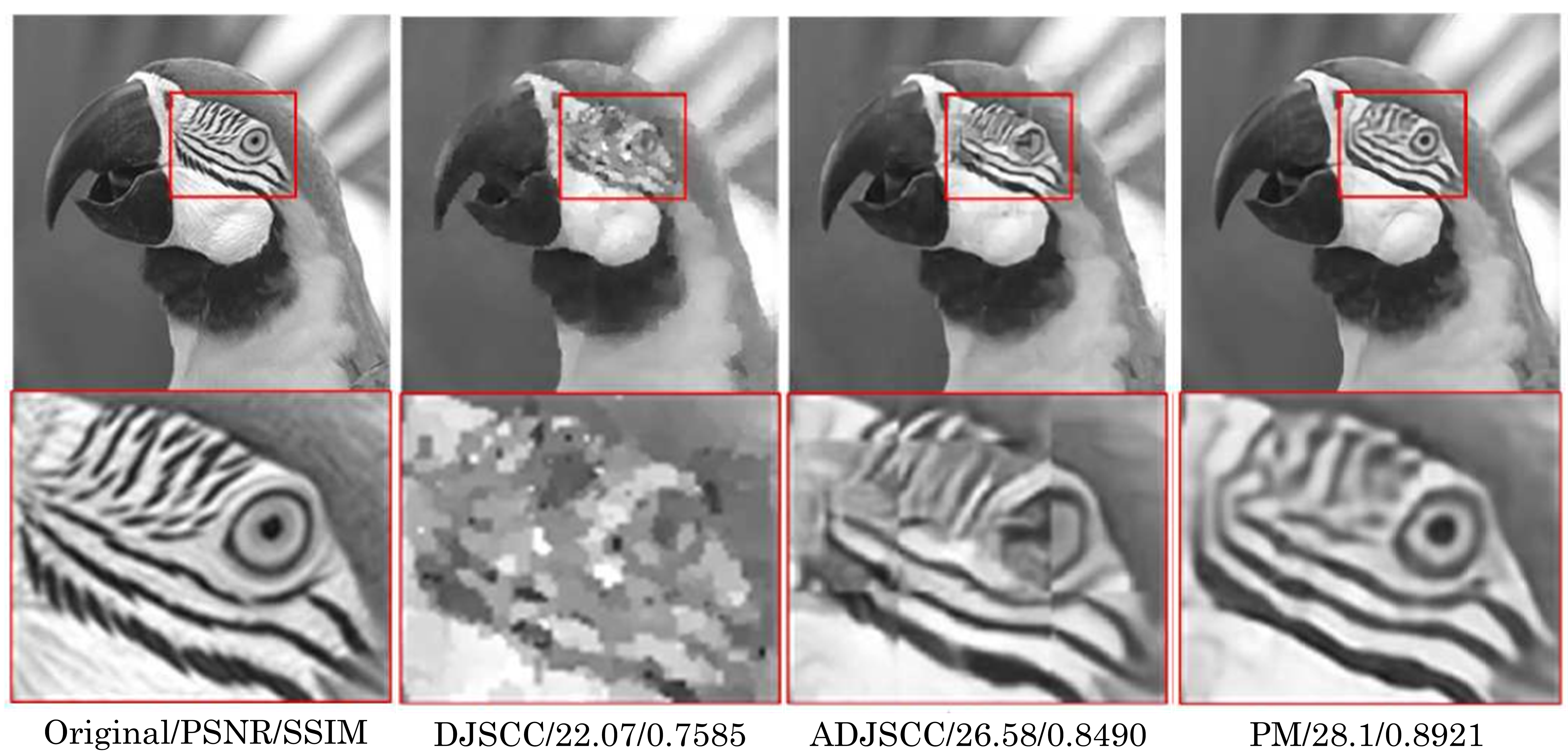}
	\caption {Example of reconstructed images produced by the proposed method with compression ratio of $1/6$ and SNR value of 4dB (PM=Proposed Method)}
	\vspace{-4mm}
\end{figure}

\subsection{Evaluation on Kodak Dataset }
The proposed approach is also evaluated using higher resolution images. To achieve this, the proposed architecture is trained on the Imagenet dataset \cite{R7}, a widely used dataset in this domain comprising around $1.2$ million images. The images are subjected to random cropping to generate patches of dimensions $224\times 224$, which are then processed in batches of $32$ samples through the network. For this set of experiments, we set $B=32$ and $l=3$. The model's learning rate is set to $10^{-4}$, and training continues until convergence is achieved. For the Imagenet dataset, the model is trained using $SNR_{\text{train}}$ values of $4dB$. This involves splitting the dataset into a $9:1$ ratio for training and validation purposes. For final assessment, the Kodak dataset is employed \cite{R19}. During the evaluation process, each image is transmitted $100$ times, allowing the performance to be averaged across multiple instances of the random channel. The evaluation scenario involves the consideration of AWGN channel.

Fig. 5 presents the comparison of average PSNR against SNR for a compression ratio of $1 / 6$ with DJSCC and ADJSCC. The results in Fig. 5 demonstrate that our method outperforms DJSCC and ADJSCC, capturing critical visual details in compressed images for better quality reconstructions. This indicates consistently higher PSNR values across varying SNR levels, highlighting improved image fidelity and minimized channel-induced distortions. The proposed technique excels in delivering superior image quality with increased compression level, even in noisy conditions.

Finally, a visual comparison of the reconstructed images for the proposed scheme trained for CIFAR-10 under consideration in AWGN channels in comparison with DJSCC and ADJSCC is presented in Fig. 6. For each reconstruction, we calculated the PSNR and structural similarity index measure (SSIM) values. Based on the given results, it is clear that the proposed method exhibits excellent visual reconstruction ability. Also, the method can more accurately restore the details of the original image. It can be concluded that the proposed method is capable of preserving the image quality and results in a superior compressing and reconstruction performance.

\section{Conclusion}
This paper introduces a novel deep joint source-channel coding algorithm for efficient image transmission over wireless channels. The approach combines block-based CS with DL techniques to design a joint source-channel encoder. This encoder employs a CNN-based model for image compression, enhancing resilience against noise by proper encoding. The model integrates an adaptive CNN-based sampling matrix to capture structural information for improved compression and encodes compressed images into complex-valued signals that adhere to the average power constraint. The decoder network, comprising CNN-based layers, reconstructs high-quality images from channel-encoded data. Through joint training, the proposed method minimizes the loss function for high-quality image reconstruction. Evaluations on CIFAR-10 and Kodak datasets highlight the method's superior performance compared to DJSCC and ADJSCC frameworks. Our approach consistently outperforms these methods in terms of PSNR across varying SNR and compression ratios, showcasing its effectiveness in achieving robust image transmission in the presence of wireless channel noise. This synergy between CS principles and DL-based techniques presents a promising solution for improved image compression and reconstruction in wireless image transmission systems.

\bibliographystyle{IEEEtran}
\bibliography{Conf-paper}

\end{document}